\documentclass[letterpaper, 10pt, times]{IEEEtran}  
\usepackage{color}
\usepackage{amsmath,amssymb}
\usepackage{diagbox}
\usepackage{algorithmic}
\usepackage{algorithm}
\usepackage{subcaption}
\usepackage{graphics}
\usepackage{float}
\usepackage{import}
\usepackage{epsfig}
\usepackage{hyperref}
\usepackage{flushend}
\usepackage[utf8]{inputenc}
\usepackage[english]{babel}
\usepackage{fancyhdr}
\usepackage{lastpage}
\newtheorem{theorem}{Proposition}
\title{Improving Blockchain scalability based on one-time cross-chain contract and gossip network}

\author{Keyang Liu, Yukio Ohsawa$^{1}$
\thanks{$^{1}$ Department of System Innovation, Graduate school of Engineering, University of Tokyo	}}

\begin{document}

\maketitle
\thispagestyle{empty}
\pagestyle{fancy}

\begin{abstract}
This study proposes a novel solution that provides secure interoperability for blockchains, which improves the overall scalability of the whole blockchain network. In our solution, a cross-chain task will build a one-time cross-blockchain contract. Each blockchain system can follow the contract to complete or this task. The result of tasks is bound with the system, hence can be anchored to all other blockchain systems through the gossip network. This work shows our result can provide linear scalability for the whole system and achieve consistency among honest systems.
\end{abstract}

\section{INTRODUCTION}

Blockchain, the solution of achieving consensus in a decentralized system, has been adopted in fields such as finance \cite{Blockfinance}, supply chain \cite{BlockSupply}, and crowd-sourcing \cite{Blockcrowd}. Currently, blockchain technology can provide two functions in a decentralized system. On one hand, blockchain can naturally work as a secure distributed ledger \cite{VTL} for storing data redundantly and correctly. On the other hand, blockchain can provide a reliable, distributed calculating platform by enabling smart contracts \cite{sc}. In a blockchain system, all nodes can execute tasks consistently and offer robust services to all users with stable quality.

\begin{figure}[H]
	\centering
	\includegraphics[width=0.5\textwidth]{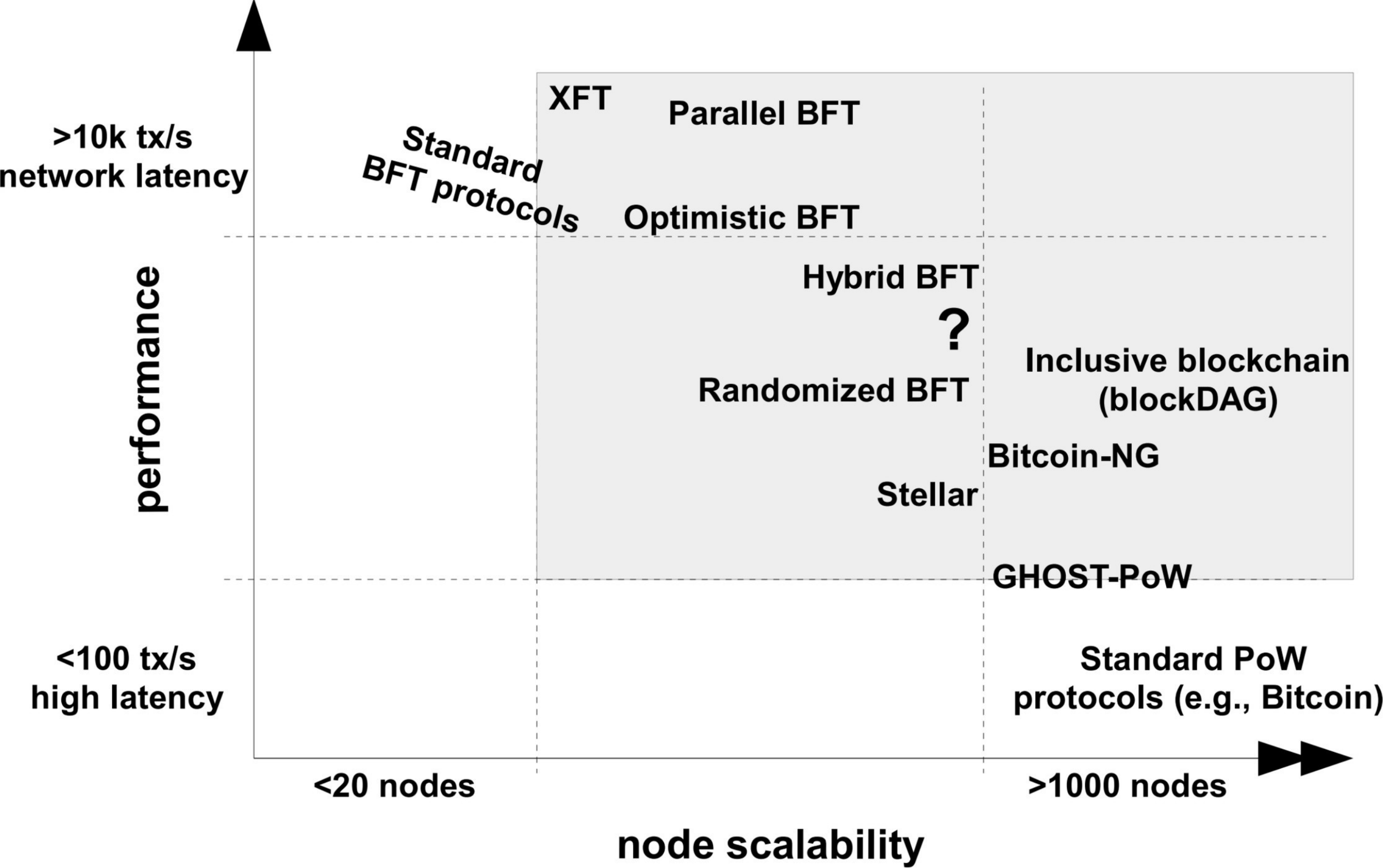}
	\caption[The scalability of blockchain systems]{The scalability of Blockchain systems \cite{vukolic2015quest}.}
	\label{fig: scalability}
\end{figure}

The advantage of a decentralized system is its robustness and trust-free setting compared to the centralized system, considering crash and Byzantine faults \cite{Byzantinefault}. In return, decentralized systems have to handle the problem of scalability instead. As Figure~\ref{fig: scalability} shows, scalability includes two evaluation dimensions: performance and node scalability. Performance usually refers to the throughput of a system, which reflects the its utility. In a standard proof of work (PoW) scheme, the frequency of producing one block is statistically near a fixed value which determines the difficulty and security. As a result, the performance of the standard PoW scheme relates to the maximum size of each block and network latency. On the other hand, node scalability shows the number of nodes a system allows in reality. In standard Byzantine fault tolerance (BFT) protocols, the communication cost is $O(n^2)$ for each round, where $n$ is the number of nodes. Hence, standard BFT protocols cannot support the case requires more than one hundred nodes.

However, if a system aims to improve performance and scalability at the same time, consistency is difficult to achieve. If one system achieves high throughput and node scalability at the same time, the system must achieve consistency between all nodes within a limited time. However, without enough time to confirm different nodes, it is difficult for nodes to tolerate Byzantine faults. In an asynchronous network, this requirement means that consistency strongly relies on a trustworthy assumption. This property leads to many consensus algorithms under different security settings \cite{karame2016security} and results in plenty of solution trade-offs between scalability, throughput, and security in the field of blockchain system design.

Side chain introduces an idea that shifts some payload of the blockchain into a quick response off-chain network for fast service \cite{back2014enabling}. The blockchain only confirms the final result of these payloads. The design of a Hash Time Locked Contract (HTLC) \cite{decker2015fast} provides an implementation of the lightning network into a cryptocurrency field, which improves the performance of standard PoW protocols. Sharding tries to split payloads into several shards for parallel processing and converge all final results into the main chain's block \cite{luu2016secure}. Sharding require each independent shards to make consensuses independently, and use cross-shard communications to protect the consistency. Finally, use a mechanism to combine those results together to achieve overall consensus for the whole blockchain system. 

A new direction for improving scalability is interoperability. Interoperability allows for multiple blockchains to cooperate even if they have different consensus rules and data structures \cite{wegner1996interoperability}. The value of interoperability is that it allows for the different blockchains to provide services independently and extend services with the help of other blockchains. Unlike sharding, different blockchains can provide different services, so each blockchain can be more independent compared to sharding. However, with the help of interoperability, the result of one blockchain system can be used to support services on other systems, which allows for many systems to provide services simultaneously. One typical use case of interoperability is the sharing of health data \cite{gordon2018blockchain} \cite{azaria2016medrec}\cite{Vora2018Health}. Interoperability also protects data privacy better compared with other methods. \cite{Cross-Chain}. In this cross-chain scenario, only trusted blockchain can access to all data of its client and user need to grant permissions for all spread of data by cross blockchain communications. However, the loose connection between different blockchains also raise the cost of communication and consensus. In next section, we will introduce more prototype of cross-chain frameworks and analysis their pros and cons. 

This study tries to improve scalability through distributing the payload into different blockchain systems, which guarantee the block is finalized after a fix number of blocks(finality), and granting interoperability to them. We will propose a framework to handle cross-blockchain communications for blockchains satisfies certain requirements. By considering the possible crash or Byzantine faults of each blockchain, this work focuses on providing secure interoperability between two blockchains and detecting any forks(more than one independent branch are valid in the same system) that can affect the other systems. 

Our main novelty are as follows: 
\begin{itemize}
	\item 1. To our best knowledge, we propose the first finality based protocol for cross-blockchain communications that can solves the delivery versus payment (DVP) problem \cite{DvP} for cross-chain tasks.
	\item 2. To our best knowledge, we are the first team to consider the possible corrupted blockchain system during cross-chain tasks, and include the defense of it into our design.
	\item 3. To our best knowledge, we are the first team to use a gossip network to detect attacks of a Byzantine fault blockchain and control the damage by configurable settings of the protocol.
	\item 4. To our best knowledge, we are the first team to adopt a dynamically adjust mechanism to handle the cost of making consensus about cross-chain tasks among different blockchains so that we can reduce the burden of synchrony different blockchains.
\end{itemize}

\section{PROBLEM CLARIFICATION}
This section will clarify the problem model on which this paper focuses and the benchmark used to evaluate our result.
\subsection{Notations}

$\mathcal{S}$ represents a network of $n$ blockchain systems, where system $S_i$ has $q_i$ nodes. Each system has a distributed network of nodes that maintains a list of blocks $BC$. $BC$ stores all finalized blocks produced according to the consensus algorithm of the system. The position of a block in the blockchain is called its height. To simplify the notation, If more than $r_i$ nodes has confirmed the block and more than $t_i$ blocks are made after this block, we say this block is finalized. $r_i$ and $t_i$ is determined by the consensus algorithm of $S_i$. e.g. $r=\frac{2}{3}q_i, t=0$ in Practical Byzantine Fault Tolerance($PBFT$) and $r=1, t=6$ in PoW.

A block in each system's blockchain contains at least two parts: header and body. A block header contains at least the hash of the previous block, metadata of the block body, and signatures of its creator. Block body is an ordered collection of transactions. If a block of $BC_i$ contains one transaction $tx$, we say $S_i$ committed $tx$(i.e $\exists b \in BC_i, tx \in b.body$). 

We assume that there is a request channel between every two systems, which means for $S_i$ and $S_j$, at least $r_i$ nodes in the system $S_i$ can send requests and obtain responses from $r_j$ nodes of the system $S_j$. As a result, $S_i$ can achieve a consensus about the state of all systems with a limited error. This request channel is essential for transporting messages between, we will review it in Section V. 

we divide the task of the whole system network into two types. First part contains local task for each system that can be verified and committed within the blockchain independently. The other part includes cross-chain tasks. For completing a cross-chain task $(tx_i,tx_j \dots tx_m)$ of system $S_1, S_2 \dots S_m$, each system $S_i$ has to commit $tx_i$. In this paper, our proposal guarantees that any cross-chain tasks among honest system will be complete without trust, and any malicious behavior will be detected. All symbols are listed in table~\ref{tab: symbols} for reference.

\subsection{Adversary Model}

We use a fixed proportion adversary model where each system $S_i$ has a portion of $f_i <= 1$ nodes is controlled by the adversary. 

This adversary model is an extension of most blockchains' security assumptions. Since each consensus algorithm has identical assumption of malicious node, each system cannot trust the finality of other systems. This assumption is reasonable especially when the network $\mathcal{S}$ is permission-less. If $S_{adv}$ tampered $BC_{adv}$, it may affect its cross-chain tasks with $S_i$ and $S_j$ if they are in different branches. This will finally consistency of local tasks in $S_i$ and $S_j$. This kind of attack is the main risk accompanying with interoperability.

\section{RELATED WORKS}
This section introduces works that are critical to understanding the problem.

Chen et al. \cite{IBC} and Kan et al. \cite{IBCA} have studied communication between different blockchains by simulating Internet stack and TCP protocol to manage inter-blockchain communication.Tendermint \cite{Tendermint} and Polkadot \cite{Polkadot} also provide some attractive designs for blockchains' interoperability.  Their products has been implemented in real markets (https://tendermint.com/, https://polkadot.network/). These works try to create a backbone blockchain whose validators represent independent blockchains to record all inter-blockchain communications. For better security, they all introduce some nodes outside of the system to monitor the behavior of the validators. Borkowski et al. \cite{DeXTT} proposes a DeXTT protocol for cross-chain token transferring. This work uses a claim messages to register the proposal of cross-chain tasks and use incentives to motivate validator reporting the hard behavior. However, as the authors admitted in \cite{TowardsSchulte2019}, this work is lack of scalability due to the independence of each cross-chain tasks. 

Omniledger \cite{omni}, RSTBP \cite{RSTBP}, FleetChain \cite{FleetChain}, Chainspace \cite{chainspace} used a 2 phase commit protocol to help cross shard communication to be accepted. They require the client to send transactions to the directly related shards in the prepare phase and then commit all these transactions in more shards to achieve the reliability of the cross-chain tasks. In our works, we also use a similar process to resolve the DvP issue. Besides, instead of using a third system to prove reliability, we let the whole network to guarantee the finality of each system.

Monoxide\cite{Monoxide} and RChain\cite{Rchain} used a relay transaction system to resolve the trust issue between two systems. These works usually commit a part of cross-chain tasks in one shard. Then, this shard will send a relay transaction to the other shard and trigger the rest part of the task. The relay transaction can only be accepted if it is confirmed by a number of blocks in the first shard. 

Another typical problem is how to build communications between two blockchains. Hardjono et al. \cite{DesignPhilosophy2018} investigate this topic and separate different designs into passive mode, which only listen to one blockchain, and active mode, where each blockchain can be a sender and receiver. One important observation they have made is the active mode need HTLC or similar mechanism to build trust. Hence, in our work, we use finality and view of each blockchain to build a gossip network. Based on the gossip network, each blockchain need to implement a minimum API as described in Section IV. Proposal A. like the work of Scheid et al. \cite{Bifrost}.

One innovation our framework has made is using a gossip network to guarantee the security of cross-chain tasks. Gossip algorithm is a simple, efficient, and robust that very suitable for P2P communications. Works like \cite{He2019Gossip}, \cite{Allombert2019Tezos}, and \cite{Berendea2020Fair} already introduces them into blockchain field.
In this works, we first use gossip network to build a continuous safety network for different blockchain systems. 

\section{PROPOSAL}

We assume all systems $S_i$ adopts various consensus algorithm that guarantee the finality and handling tasks requested by clients. All the systems are connected by gossip network (subsection~\ref{sec:gossip}) and request channel(subsection~\ref{sec:communication}). Whenever a cross chain tasks is received, all related system will run a cross blockchain contract(CBC) (subsection~\ref{sec:CBC}) upon the request channels to complete the tasks. The gossip network will keep running whenever there is a CBC or not and provide the security of CBC.

\subsection{Functions}
This section introduces several functions which are necessary and predefined for each system as the foundation of our first contribution: Cross Blockchain Contract.

First, each node of a system $S_i$ maintains a view list for views of all other systems in the network that system $S_i$ can access. View of $S_i$ contains the aggregated $BC_i$, latest $k$ finalized blocks' Hash, as well as sufficient proofs (e.g., $r_i$s' signatures or threshold signature from the system).A system can generate its view according to Algorithm~\ref{alg: JHash}. Here, $H()$ is a universal Hash function, and $GenProof()$ returns the proofs.
\begin{algorithm}
	\caption{View generation}
	\label{alg: JHash}
	\begin{algorithmic}[1]
		\renewcommand{\algorithmicrequire}{\textbf{Input:}}
		\renewcommand{\algorithmicensure}{\textbf{Output:}}
		\REQUIRE Blockchain $BC$, $k$
		\ENSURE  view of the system
		\STATE{$V=[]$}
		\STATE{$m= BC.Length$}
		\FOR{$i$ in [0,m-k] }
		\STATE{$HeaderHash=H(BC[i].header)$}
		\STATE{$V[0]=H(Result[0]||HeaderH)$}
		\ENDFOR
		\FOR{$i$ in [m-k,m-1] }
		\STATE{$HeaderHash=H(BC[i].header)$}
		\STATE{$V.append(HeaderHash)$}
		\ENDFOR
		\STATE{$V.append(GenProof(Result[K]))$}
		\RETURN V
	\end{algorithmic}
\end{algorithm}

Each system needs to maintain $O(n*(k+1))$ hash values as a local view lists(roughly $O(n*(k+1))$ kilobyte). This copy is cache of the whole network's latest status.  If $S_i$ sends its view lists to $S_j$, the view of $S_k$ in $S_i$'s view list, denoted by $V_{ik}$, can update $V_{jk}$ as long as $V_{ik}[0]$ can be calculated from $V_{jk}$. Hence, whenever a message that contains view information is received, the node will use this message to update their existed view list without confirming $S_k$. Each system broadcast this message in the gossip network to keep catching up the state of others.

Second, each system should implement two public query method: $verify(tx)$ and $check(hash)$. $verify(tx)$ will inform whether the tx can be added to the next block. e.g. In Bitcoin, the input should be a subset of unspent transaction output (UTXO), and the sum of inputs should be no less than the sum of outputs. $check(H(tx))$ returns the position of one tx in $BC[]$ of $-1$ if it is not existed. Formally speaking, for any system $i$, if $\exists h$ s.t.  $tx \in BC[h], check(tx)=h$; otherwise, $check(tx)=-1$. Hence, the following properties hold:

\begin{itemize}
	\item a. (Validity) For any blockchain with maximum height $h$, if $verify(tx) == True$, $check(H(tx))==-1$.
	\item b. (Agreement) For any system $S_i$, if $check(H(tx))>0$, $tx$ can be accessed from at least $r_i$ nodes.
\end{itemize}

The $check()$ function allows other systems to check the existence of a $tx$ in the $BC$. Each honest node has the responsibility to reply the request of $check()$ from any sources. When a system sends or replies a $check()$ request, it will send the result attached by its view.

Third, we define a contract transaction for system $S_i$ and $S_j$ as: $$ctx_i:=H(tx_i)||H(tx_i')||h_i||j||H(tx_j)||H(tx_j')||h_j$$
$tx_i$ is the cross-chain task's part for $S_i$, while $tx_j$ is the rest part in $S_j$. $tx_i'$ is the reversed transaction of $tx_i$. For example, if $tx_i$ send coin from $A$ to $B$, $tx_i'$ transfer the same amount of coin from $B$ to $A$. The expiration height for this condition transaction is $h_i$ in system $S_i$ and $h_j$ in $S_j$. The meaning of this contract transaction is as follows: $tx_i$ and $tx_j$ should be verified before height $h_i$ and $h_j$ in the respective blockchain, otherwise, system will commit $tx_i'$ and $tx_j'$. Similarly, we can create a symmetric contract transaction for system $S_j$ as $$ctx_j:=H(tx_j)||H(tx_j')||h_j||i||H(tx_i)||H(tx_i')||h_i$$ By definition, $verify(ctx_i)=verify(tx_i)$.

Fourth, we define $tx \rightarrow  tx'$ represents $tx$ is a precondition of $tx'$ (e.g. $tx_i\rightarrow  tx_i'$).A transaction $tx$ is locked means for any $tx^*$ that $tx \rightarrow tx^*$, $verify(tx^*)==False$. A system keeps all its committed contract transactions in a waiting list before they are expired. Whenever a contract transaction $ctx$ is in the waiting list, $\forall tx$ s.t $ctx \rightarrow tx$, tx is locked.

\subsection{Cross Blockchain Contract(CBC)}
\label{sec:CBC}
CBC is used to help several independent system to reach the atomicity of committing all transactions together. This protocol mainly make use of the finality of each system so each single system cannot reverse their commitments unless some systems did not commit the transaction. The CBC contains 3-phase committing: 1. committing the $ctx$ meta data, 2. commit $tx$ and lock the result, if all systems has committed the $ctx$ 3. if the step 2 failed, then commit the reverse function $tx'$. The whole protocol relies on the finality of each system and only the systems that need to commit transactions will directly be affected.

The CBC works as following algorithm~\ref{alg: CBC}:
\begin{algorithm}
	\caption{Cross Blockchain Contract}
	\label{alg: CBC}
	\begin{algorithmic}[1]
		\renewcommand{\algorithmicrequire}{\textbf{Input:}}
		\REQUIRE Blockchain $ctx_i$, $tx_i$, $tx_i'$
		\STATE{$H(tx_i)||H(tx_i')||h_i||j||H(tx_j)||H(tx_j')||h_j = ctx_i$}
		\STATE{$ctx_j =H(tx_j)||H(tx_j')||h_j||i||H(tx_i)||H(tx_i')||h_i$}
		\IF{$Verify(ctx_i)==false$}
		\STATE{Abort}
		\ENDIF
		\STATE{$Commit(ctx_i)$}
		\WHILE{height < $h_i$}
		\STATE{$Result = Check_j(ctx_j)$}
		\IF{$Result > 0$}
		\IF{$Verify(tx_i)==true$}
		\STATE{$Commit(tx_i)$}
		\ENDIF
		\STATE{Leave loop}
		\ENDIF
		\STATE{Wait for a while}
		\ENDWHILE
		\STATE{Wait till $S_j$ reach $h_j$ in view list}
		
		\STATE{$Results = (Check_j(H(tx_j)),Check_j(H(tx_j')))$}
		\STATE{TimeOut($Results$)}
		\RETURN
	\end{algorithmic}
\end{algorithm}

The timeout process is described in Algorithm~\ref{alg: Timeout}.
\begin{algorithm}
	\caption{Function of Timeout}
	\label{alg: Timeout}
	\begin{algorithmic}[1]
		\renewcommand{\algorithmicrequire}{\textbf{Input:}}
		\renewcommand{\algorithmicensure}{\textbf{Output:}}
		\REQUIRE $ctx$; $tx_i'$; $hc$ :=$check(H(tx_j))$; $hc'$ := $check(H(tx_j'))$
		\ENSURE  None
		\STATE{Enter Lock}
		\STATE{remove $ctx_i$ from waiting list}
		\STATE{$h=check(Hash(tx_i))$}
		\IF{$h<0$}
		\RETURN
		\ENDIF		
		\IF {$hc<0$ or $hc>ctx[5]$ or $hc'>0$ }
		\STATE{$commit(tx_i')$ and attach proofs}
		\RETURN
		\ENDIF
		\STATE{Leave Lock}
	\end{algorithmic}
\end{algorithm}

Timeout is used to unlock $tx_i$ and finish the contract. Timeout will always be executed when both expired times are reached. The correctness of this scheme is proven in the next section. Since cross-chain tasks are independent to the local task, such a workflow will increase the latency of cross-chain tasks. If the one system $s_i$ commits $tx_i$ after $h_i$, this behavior it self is a proof of the byzantine systems.

\begin{figure}[h]
	\includegraphics[width=0.5\textwidth]{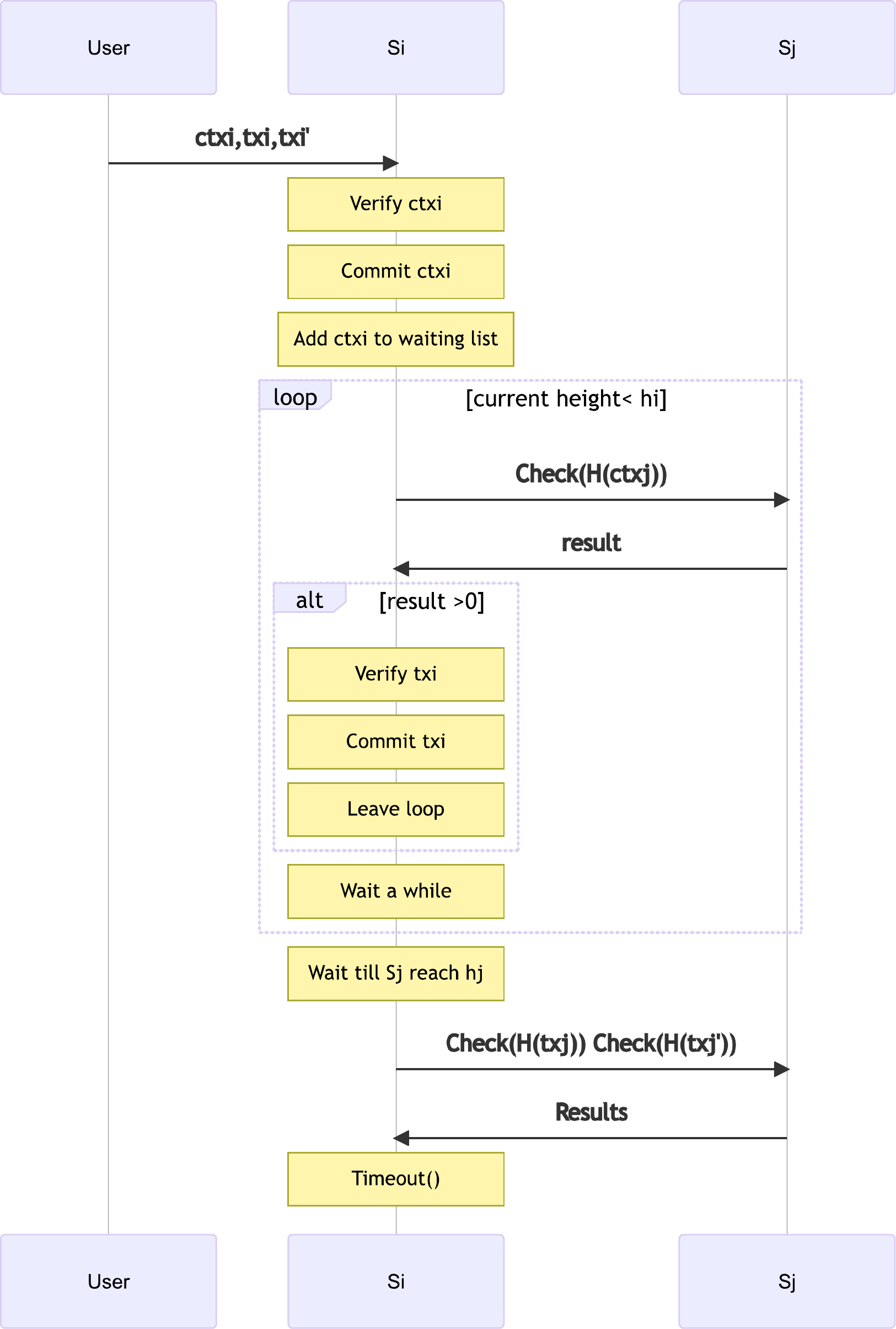}
	\caption{Cross Blockchain Contract.}
	\label{fig: workflow2}
\end{figure}

The success case of the workflow is shown in Figure~\ref{fig: workflow2}.

We can easily extend above CBC to multiple blockchain by including more components into the contract transactions. Hence, we only focus on two system cases in later analysis.

\subsection{Blockchain-wise Gossip Network}
\label{sec:gossip}
We have clarified the main steps of handling cross-chain tasks and have a sense of the cost of availability. But the CBC are fully based on the finality of systems, which is not always reliable in a blockchain network. Our second contribution is using a gossip network to guarantee the reliability of cross-chain tasks.

\begin{figure}[H]
	\centering
	\includegraphics[width=0.5\textwidth]{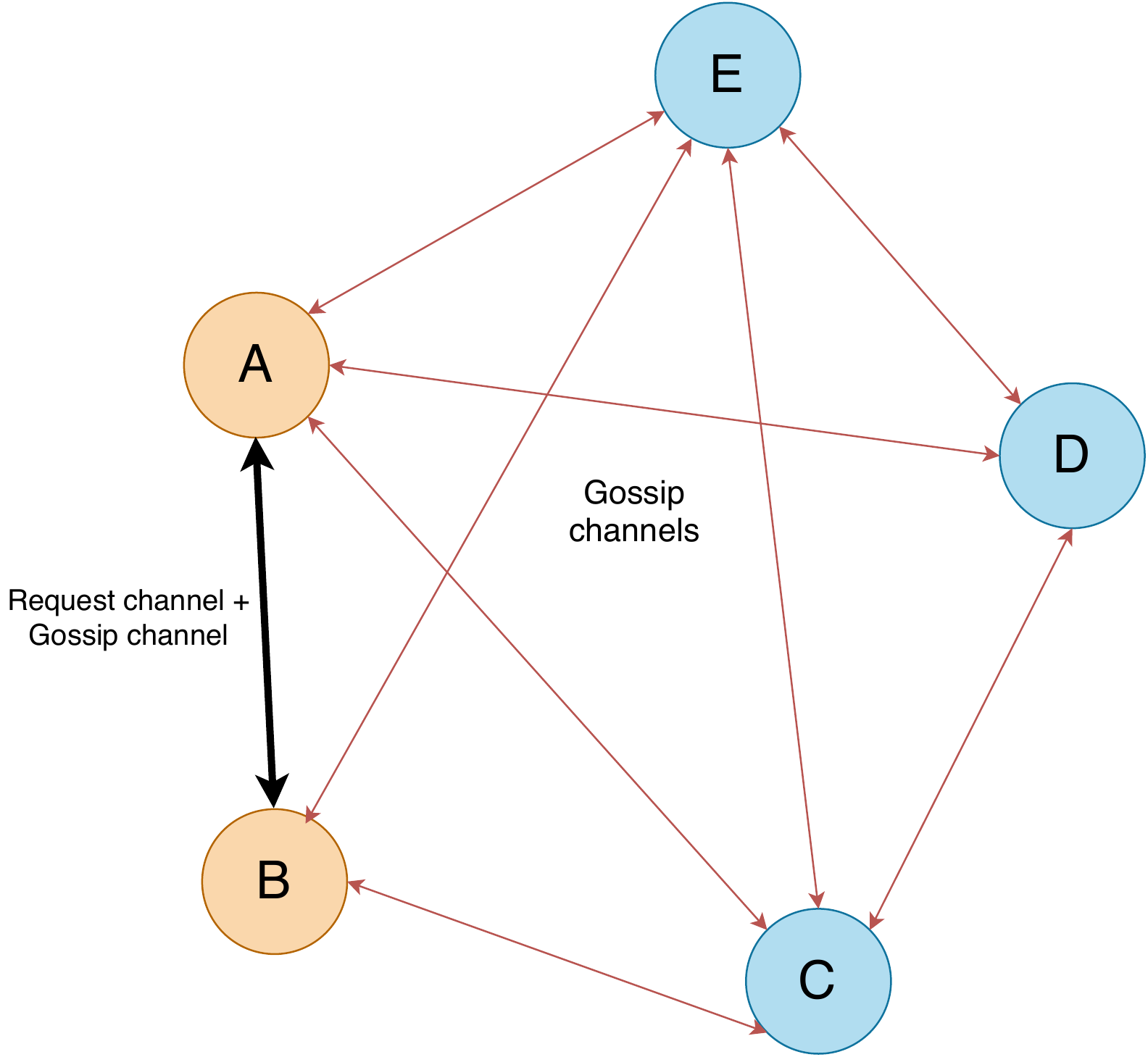}
	\caption[Network Structure]{An example network topology of systems}
	\label{fig: network}
\end{figure}

Fig~\ref{fig: network} shows an example of network topology. \textbf{A} - \textbf{E} are blockchain systems within a gossip network for passing gossip messages. A gossip channel between two systems can be used to send gossip messages. A system may connect to all other systems (like \textbf{A}, \textbf{C}, and \textbf{E}) or only link to a limited number of them (\textbf{B} and \textbf{D}). We adopt a naive gossip protocol that, whenever a cross-chain task related transaction is committed, the system will broadcast its latest view list in the gossip network, it will send the view list to each system $S_i$ with a fix probability to avoid flooding messages. The gossip protocol will guarantee any information sent into the system will be received by all systems eventually, if the system is not isolated by Byzantine systems. Request channels only needs to be build between systems that need to handle cross-chain tasks. If two systems has a Request channel, they always in the same gossip network. 

Gossip network has push mode(sender select receiver), pull mode(receiver select sender) or the mix. We adopt the mix mode in our design for satisfying different purposes. When a node sends or receives a request that will update its view list, it will compare the received updated view lists to its local copy. If the local view list needs to be updated, it will update and then broadcast the newest view list through the gossip network. If $S_i$ receives a view list from $S_j$, and $V_{jk}$ cannot be calculated from $V_{ik}$, $S_i$ needs to use pull mode to ask $S_k$ for missing info between $V_{ik}$ and $V_{jk}$ to update the view list correctly.

If any system receive a view list that conflict with its, there must be some systems behaving improperly. Hence, the system will broadcast the evidence to all systems. So all systems can reject the ongoing contract transactions with the malicious system and continue cross-chain tasks when the system resolve the conflict without breaking any existed cross-chain tasks. The conflict resolving is another topic we will not discuss in this work.

From above design, we can find the speed of spreading latest view is critical to the security of cross-chain tasks. As a result, CBC keep $ctx$ in waiting list until $h1$ and $h2$ are both reached, so the view list can be generated and spread to other systems. Each system can set a timer to push message in the gossip network to trigger an updating process.

The gossip network allows all systems to dynamically balance the security and performance. With more depth view list or increase the possibility of gossiping messages, the whole network can sync more information of all other systems hence increase robustness against byzantine system. On the other hand, by decrease the view list size and gossiping possibility, all systems can reduce the robustness for better performance.

Compare to most sharding or interoperability solutions, this gossip network is the most important difference that remove the hierarchy architecture above other shards or blockchain systems. With this gossip network, the cost of securing each cross-chain tasks is determined by the gossip network, while in most solutions, it needs independent verifier to check the context and make additional consensus.

\subsection{Communication between systems}
\label{sec:communication}
In most systems, the execution logic is handled by nodes. Hence it is important to have a brief discussion about how to implement communication between systems.

For request channel if it is permissionless setting($r =1, t>1$), the system needs to request the latest blocks as a proof. For the permissioned setting, the naive way is a full connection between all nodes of the two systems. All nodes in system $S_i$ need to send $Msg$ to all nodes in systems $S_j$ until system $S_j$'s at least $r_j$ nodes received a $r_i$ identical copies of $Msg$ for confirmation. The cost is $2r_ir_j$ times node-to-node communications for one-time request. Additionally, only one nodes need to send the request if the result is properly signed and nodes in $S_j$ can be recognized in $S_i$. This selected node will send requests to the nodes in the system $S_j$ and distribute the received results to peers in system $S_i$. Hence, the minimum times of node-to-node communications becomes $r_j$. On the other hand, if all nodes in $S_i$ can connects to all nodes in $S_j$, the communication times can be further decreased. For a given negligible probability $p_i$, a node only needs to see random $m_j$ confirmations without conflicts to believe that all honest nodes have confirmed such a result. $m$ should satisfy $(C_{r_j}^{m_j}/C_{q_j}^{m_j})<p_i$. Thus, the minimum required result can be decreased to $min(r_j, m_jr_i)$ in a permissioned network. 

On the other hand,The design of the view list can affect the performance of the gossip network. A gossip network has a fixed cost decided by $k$ in \ref{alg: JHash}. For the gossip network, each message is sending to a subset of all systems. The maximum gap of a view's height between the received message and local view lists for any system in the gossip network is set to $k$, which is used in Algorithm~\ref{alg: JHash}. $k$ affects the size of the view list and eventually decide the latency of the network. If a system cannot verify whether the received view list is correct frequently, this system should consider increasing $k$ or establishing more gossip channel between systems. For a system, each node can broadcast the gossip message within the system by internal channel, which means each node can be treated as the system in the gossip network. Hence, each system $S_i$ just needs to randomly pick $m_j$ nodes in $S_j$ for sharing the messages. Each system $S_i$ can randomly pick sender from all nodes as a source of randomness in the gossip protocol within their internal consensus algorithm.

\section{ANALYSIS}
This section provides the correctness and security proof of our method.
\subsection{Correctness proof}

The function of this CBC is to make sure the commitment of $tx_i$ and $tx_j$ are related. The CBC split this process into two steps: first, commit and lock $tx_i$ and $tx_j$. Second, check the commitment and, according to the result, decide whether to commit $tx_i'$ and $tx_j'$.

\begin{theorem}
	Assume that systems $S_i$ and $S_j$ need to accomplish the contract $\lbrace(tx_i,tx_i'),(tx_j,tx_j')\rbrace$. $S_i$ and $S_j$ must commit $ctx_i$ and $ctx_j$ before the expiration time.
\end{theorem}
\begin{IEEEproof}
If $S_i$ directly commits $tx_i$, nothing can guarantee the $S_j$ will commit $tx_j$, so $S_i$ should submit $ctx_i$ first to lock for future submission of $tx_i$. If $S_i$ commits $ctx_i$, $S_j$ have to commit $ctx_j$ so that $S_i$ will commit $tx_i$. If submission of $ctx_i$ and $ctx_j$ is after expiration time, the $Verify()$ will return false. Thus, $S_i$ and $S_j$ must commit $ctx_i$ and $ctx_j$ before the expiration time.
\end{IEEEproof}

After both systems committed the contract transaction, they will start to commit the target transaction $tx_i$ and $tx_j$ and lock them immediately.
\begin{theorem}
Timeout function finalized the result of cross-chain task.
\end{theorem}

\begin{IEEEproof}

Assume $S_i$ strictly follow CBC process. When time out happens, there will be three cases: 1.Only $tx_i$ is committed; 2. $tx_i, tx_j$ are committed; 3. $tx_i, tx_j, tx_j'$ are committed; for case 1 and case 3, line 8 in Alg~\ref{alg: Timeout} guarantees the $tx_i'$ will be committed, hence nothing else can be done. For case 2, the contract can be proofed completes, and $ctx_i$ has been removed from waiting list, $tx_i$'s result can be consumed after timeout, the cross-chain task accomplished. The same process applies to $S_j$.
\end{IEEEproof}

As a conclusion, this protocol guarantees the following conditions. 1. $ctx_i$ and $ctx_j$ are committed to the respective terminated blockchain before committing $tx_i$ and $tx_j$. 2. $tx_i$ and $tx_j$, if committed, will be locked before expiration. 3. After both systems reach the expiration time, the final result of the contract will be determined, and remove the lock of $tx_i$ and $tx_j$. If contract failed, each system will commit the necessary reversed transactions to close the system. For two honest systems, they will commit at most 4 transactions in both the successful case (e.g., $ctx_i$, $ctx_j$, $tx_i$, and $tx_j$) and the failed case (e.g., $ctx_i$, $ctx_j$, $tx_i$, and $tx_i'$)

\subsection{Security Analysis}

This section provides some brief proofs of security for security concerns. Since all cross-chain tasks are happened between two blockchains, We assume there are two kinds of attacker $Adv1$ and $Adv2$ and have different powers for $S_i$ and $S_j$.

\begin{itemize}
	\item (Honest vs Honest)$Adv1$ controls nodes in both $S_i$ and $S_j$, but cannot finalize any malicious blocks in either $S_i$ or $S_j$.
	\item (Honest vs Malicious)$Adv2$ can finalize malicious block in $S_j$ while only control some nodes in $S_i$ that cannot finalize malicious blocks. 
\end{itemize}

$Adv1$ cannot commit any invalid block in $S_i$ and $S_j$, so the only attacks they can apply is to hijack the request channel between $S_i$ and $S_j$ and send incorrect responses.
\begin{theorem}
	$Adv1$ has negligible probability to prevent honest nodes in $S_i$ and $S_j$ to receive correct message from each other.
\end{theorem}
\begin{IEEEproof}
	
	For request channel, assume $S_i$ send request to $S_j$. If $S_j$ is permissionless blockchain, the result should be attached with necessary proof (e.g. proof path in merkle tree) which cannot be provided by $Adv1$(otherwise, it can create malicious blocks). Hence let's consider the permissioned case.Since $Adv1$ cannot make blocks, it controls at most $r_i-1$ nodes in $S_i$ and $r_j -1$ in $S_j$. Hence, he cannot create either $r_j$ valid signature. If $m_jr_i<r_j$ then it must guarantees at lest one honest node in $S_i$ will only query the controlled node which has only negligible possibility $p$. Hence, Honest node can still receive correct response.
\end{IEEEproof}

On the other hand, $Adv2$ can finalize malicious block in $S_j$ hence it can easily pursue $S_i$ to accept the view $V_{ij}$ from $S_j$.

\begin{theorem}
$S_j$ cannot complete two identical cross-chain tasks on different branches $V_{ij}$ and $V_{jj}$ where $V_{ij}$ conflicts with $V_{jj}$.
\end{theorem}
\begin{IEEEproof}
For $S_i$, $V_{ij}$ is a valid view received from $S_j$, hence it will store the view in its view list. Since $Adv2$ only controls a part of nodes in $S_i$, it cannot prevent honest nodes in $S_i$ to spread the view list to other systems before $h_i$ is reached. Hence, $Adv2$ cannot prevent other systems to accept $V_{ij}$ as the view of $S_j$. If $S_j$ spread a view $V_{jj}$ that $V_{jj}$ conflicts with $V_{ij}$ and start a new cross-chain task with $S_k$, $S_i$ and $S_j$ will spread the two conflict view in the gossip network. As long as $h_i$ and $h_k$ is well adjusted according to the network,  then at least one system will find out $S_j$ breaks the finality promise and at least one of $S_i$ and $S_k$ can cancel the task before reaching $h_i$ or $h_k$.
\end{IEEEproof}

This proposition shows the importance of gossip network for preventing conflict branch attacks. Another possible attack is $S_j$ returns forged $Check()$ result that is not true for $V_{ij}$. To prevent this attack, the proof for each claim is very important, which is another topic for blockchain technology. $S_j$ can also commit $tx_j'$ after completing the cross-chain task. However, as the contract transaction is completed, this obvious attack can be easily proof by the $BC_i$ and $BC_j$. 

\section{EVALUATION}
In this section we mainly evaluate the performance of the proposed system. Below simulations did not consider the attack vectors because any conflicts will freeze the CBC with the blockchain that shows conflicts. each message between the systems is count as 1 message while indeed it should contains a group of messages from different node of the system according to section~\ref{sec:communication}.
 
\subsection{Benchmarks}

We evaluate our result from two aspects: 1. The cost of availability and 2. The cost of security. 

We use the number of cross-chain message per cross-chain task to evaluate the cost of availability. This message flows in the request channel and reflects the resources spent for cross-chain tasks, hence it can be used to show the cost of availability. For security, the gossip network is mainly used for the consistency of each systems. The average message on the gossip network directly reflects the cost of the gossip network. For different security levels and scales, the change in such cost can reflect the scalability for security guarantees. If nodes needs to send more data in gossip network, the communication between different blockchains will become more frequently so the attacks could be easily checked, the whole system is more secure. On the contrary, if fewer messages are sent through gossip network, one Byzantine blockchain could have higher possibility to foolish other system

In the following parts of the paper, we provides a simulation experiment based on python to evaluated the average gap of view lists and the maximum gossip messages and request message ($check()$) to different network scale and cross-chain tasks rate. Second, we will compare our results with other solutions of interoperability to highlight pros and cons.

\subsection{Performance Simulation}
Our experiment simulates the number of messages between different systems within network and proof that our system can achieve a linear scalability. The result only contains message in the request channel. the intra-blockchain communication depends on the consensus algorithm of each system, and usually linear to the number of transactions to commit which is stable in all settings of the experiments. To our best knowledge, there has been no work trying to achieve same goal in this form, we believe our work can provide a good direction for the study of blockchain systems' scalability. All parameters of this simulation are listed in Table~\ref{tab: pars}. The setting parameter includes three digits: the scale of the system ($n$), rate of cross-chain tasks for a block($p_c$), and rate of sending a gossip message to a system(In push mode, the possibility of sending view list to another system). The number of cross-chain task in a block follows a geometric distribution, that averagely start $\frac{p_c}{1-p_c}$ cross-chain tasks in a block. All these cross-chain tasks are send randomly to another system. For example, 3\_1\_1 represents a network with 3 nodes, 1 of 9 blocks contains a contract transaction, a gossip message will be sent to each system with a possibility of 10\%. The network topology forms a strongly connected graph. The gossip network is simulated by a broadcast channel to follow the rule of the gossip algorithm. 
\begin{table}[]
	\begin{tabular}{|p{0.65\linewidth}|l|}
		\hline
		Parameter name & value\\\hline
		block creating speed  & 0.1s/block \\\hline
		number of block & 10,000 \\\hline
		the scale of the system ($n$) & 3/5/10 \\\hline
		rate of cross-chain tasks for a block  & 10\%/20\%/40\%\\\hline
		rate of sending a gossip message to a system & 10\%/20\%/30\%\\\hline
	\end{tabular}
	\caption{The parameter used in the simulation experiment.}
	\label{tab: pars}
\end{table}

As previous analyses have shown, the function of our framework relies on the request of the request channel. The security of our work is based on messages of the gossip network to control the gap of view. Hence, we evaluate the effects of a change in parameters to evaluate scalability. With a lower gap in view, the system will be more secure, while fewer request and gossip messages will reduce the cost of this framework.

\begin{figure}[h]
	\includegraphics[width=0.5\textwidth]{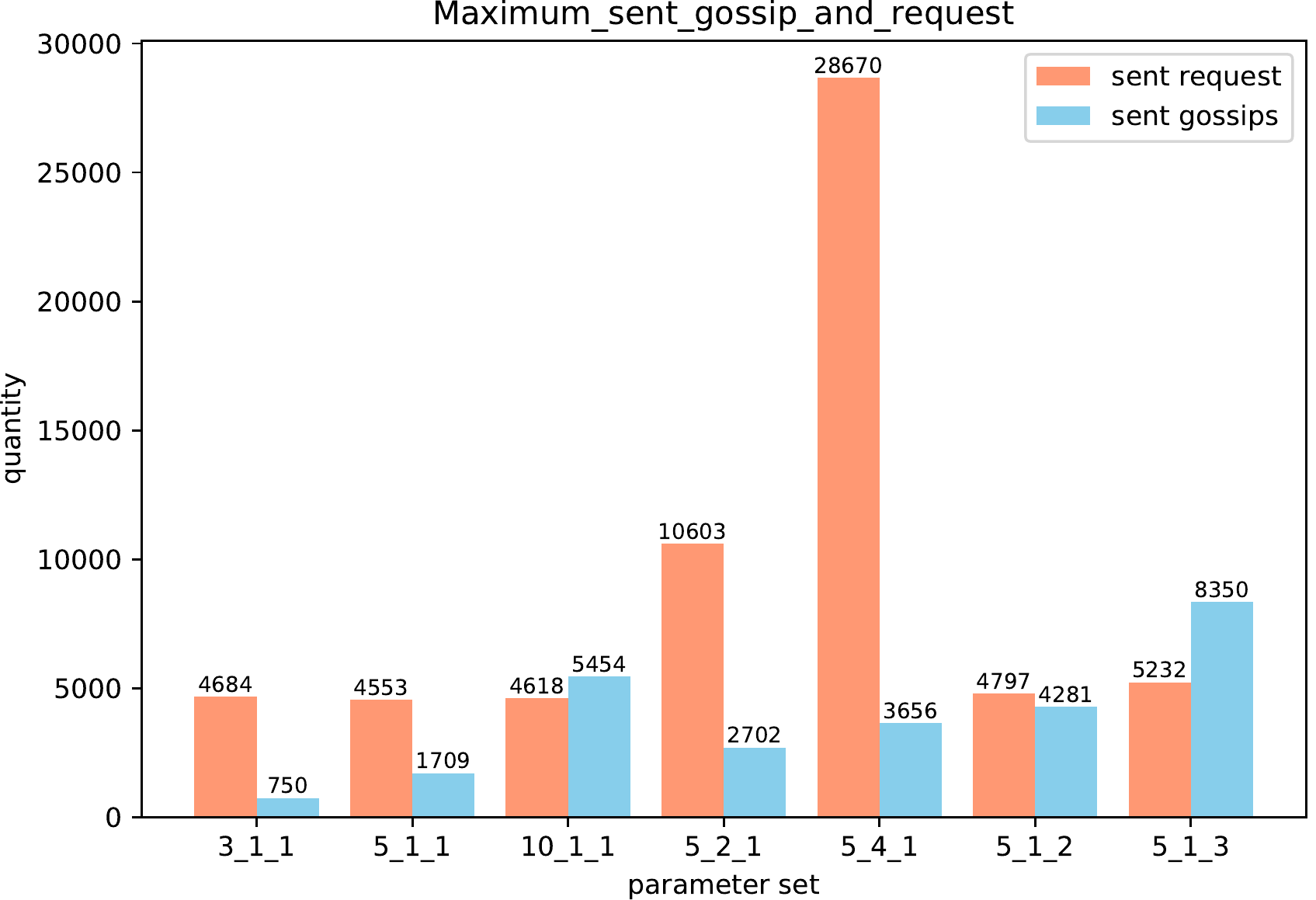}
	\caption[Sent requests and gossips]{The graph of requests and gossips sent by a system, comparing the change in scale, cross-chain task rate and rate of sending a gossip.}
	\label{fig: GRs}
\end{figure}
Figure~\ref{fig: GRs} shows the result of sent requests that are related to cross-chain tasks. Generally, one successful cross-chain task requires at least four cross-chain requests. There will be some resent happens due to network latency, but the result basically shows the message of requests is linear to the number of cross-chain tasks(1/9;1/4;2/3 compare to 4553;10603;28670).
 
For the gossip network, the scale will cause a significant change to the required gossip messages number that is growing faster than a linear function(3:5:10 vs 750:1709:5454). The rate of cross-chain tasks per block, which leads to more view updates messages, only increases sub-linearly. This is because more than one cross-chain task related transactions' commitment will only trigger one gossip message. Finally, increasing the rate at which gossip messages are sent will significantly increase the volume of gossip messages, which is trivial. 
\begin{figure}[h]
	\includegraphics[width=0.5\textwidth]{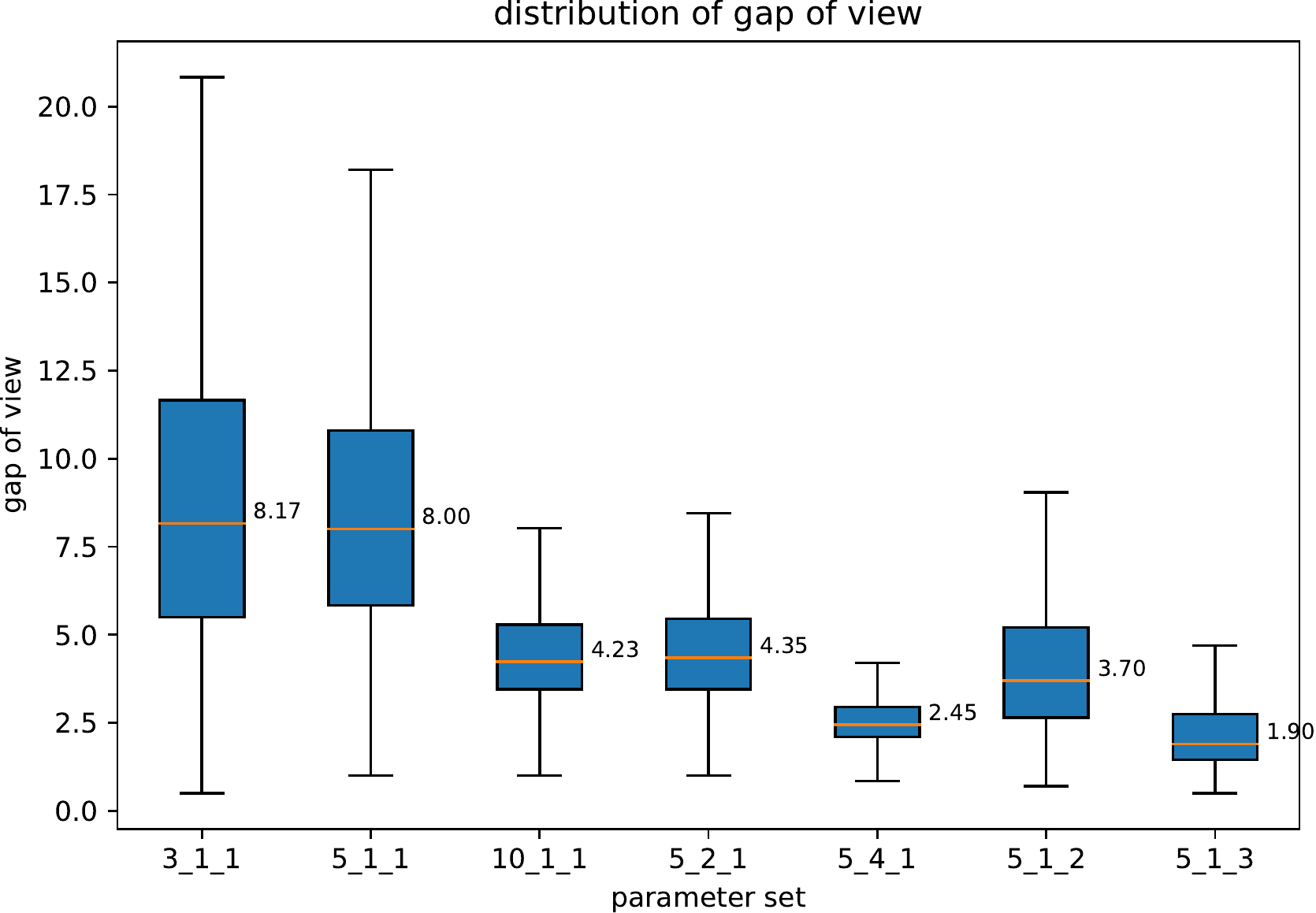}
	\caption[Gap of views]{A distribution of the gap of views for systems, comparing the change of scale, cross-chain task rate and rate of sending a gossip.}
	\label{fig: Dgap}
\end{figure}

Figure~\ref{fig: Dgap} reflects the distribution of gaps between view lists and the real states. According to the result, we can find that the gap decreased as the scale increased, which is the direct result of more gossip message in the gossip network. One impressive result is the increase in cross-chain tasks results in the expectation but also the variance of a gap decrease. The reason is that each cross-chain task will broadcast more up-to-date information than a forward triggered gossip message. Finally, the gap will not disappear even we sent a lot of gossip messages, hence the choose of expiration time $h_i$ and view size $k$ is very important.

By comparing Figure~\ref{fig: GRs} and Figure~\ref{fig: Dgap}, we can find that even the gossip message send rate is less than 50\% for each system, the average gap of view can be limited. This result is useful for determining $k$ for view, view list, and the length of expiration of the system. In the simulation result, if a system increases the rate of sending gossip messages such as sending on average 1.2 systems each time (sending rate is 30\%), with less than 1 cross-chain task per 9 blocks (5\_1\_3 case), the expected gap will be less than 3, and $k$ can be set to 5 for the worst case. If more cross-chain tasks are required, $k$ can be decreased further.

\subsection{Comparisons}

In this part we will compare our works with Cosmos, Polkadot and DeXTT. All these three solutions implement an interoperability to blockchains. We choose Cosmos and Polkadot since they are mature products for interoperability, and DeXTT as another try of decentralized interoperability(no main blockchain). 

Since different proposals have identical security assumptions, it is hard to say which solution is better than the others. Hence, we will evaluate the cost of availability and security mechanism for the cross chain task mentioned in Problem Model. The cost are evaluated as the number of inter blockchain communications(communication cost) and commits of blockchains(consensus cost). Table~\ref{tab: Interoperability} list the main differences between our 4 solutions. Lastly, we want to compare the scalability of each system.

\begin{table}[]
	\begin{tabular}{|m{0.3\linewidth}|m{0.3\linewidth}|m{0.3\linewidth}|}
		\hline
		Solution & Cost of availability & Security mechanism\\\hline
		Cosmos \cite{Cosmos}  & 4 hub txs + 2 internal tx & Hub chain \\\hline
		Polkadot \cite{Polkadot} & 1 Relay Chain tx + 2 internal txs + 1 proof query & Relay Chain \\\hline
		DeXTT \cite{DeXTT} & 2 broadcast txs + 2 internal txs & Veto process \\\hline
		This work  & 2+ proof queries + 4 internal txs & Gossip network \\\hline
	\end{tabular}
	\caption{The main design differences between selected solutions}
	\label{tab: Interoperability}
\end{table}
For Cost of availability, Cosmos and Polkadot use a main chain (hub and Relay chain) to maintain the consensus between different systems. This choice is reasonable since they are building ecosystem of their products. A main chain can reduce the cost of making standard and governance. Cosmos propose an Inter Blockchain Communication(IBC) protocol to allow one system use two transactions to commit to hub and the hub needs another two commits to the destination system for proving. Polkadot use a more centralized way that commit all blocks of systems to the Relay chain as a universal proof for systems. Polkadot allows different systems making chain independently, the finality can only be confirmed after the relay chain accept the commit. Polkadot is more centralized compare to Cosmos in terms of finality. The usage of a centralized main chain to keep consistency allows them to spend shorter time(consensus making is faster than expiration time) and less resources (main chain is always finalized) to finish cross-chain task. As a decentralized solution for token transfer, DeXTT broadcasts contest transaction and finalize transaction to all other blockchains as an anchor for the task. Our framework uses 4 internal transactions and at least 2 proof queries. If each system can provide a reliable query oracles, our frameworks spent the minimum result compare to others. Our work and DeXTT all use the idea of anchoring to other system, but obviously, our systems gains more scalability since we don't need to broadcast the cross-chain tasks to all systems.

In addition to the availability of cross-chain tasks, the security of these tasks is also very important. For Polkadot and Cosmos, they need to make sure sub-systems will not reverse the result anchored in the main chain. Cosmos hub and Polkadot Relay Chain all ask the node of the main chain to check the security of each sub systems. In addition, Polkadot introduce fisherman to hunt invalid behaviors for economical incentives. These are practical but centralized method. However, they cannot guarantee security if collusion within main chain happened. This is because the consensus algorithm of the main chain still relies needs to follow the byzantine assumption. However, in our system, the security is guarantee by the finality and the gossip network, which can easily detect the byzantine behavior and prevents the attack by rejecting txs on the compromised branch of the byzantine systems. DeXTT use a similar veto systems to encourage other systems to report the double spending cross-chain tasks and use incentives to motivate them. On the other hand, our framework connect the cross-chain tasks with the finality of each system and use a gossip network to combine the consistency of all systems together. The security of our system completely rely on the gossip networks which also provide the more flexibility to each system. 

Lastly, we want to evaluate the scalability. DeXTT is not designed for scalability so its protocol will become very costly once the number of systems increases. As a return, the security is improved because more system means strict veto process. Polkadot and Cosmos's scalability depend on the consensus algorithm of the main chain. Tendermint limit the scale of hubs. On the other hand, Polkadot is more efficient on building relay chain (commit chains instead of blocks), but cannot completely resolve the scalability issue. Our framework achieve a good scalability that the cost of cross-chain tasks will not increase as the scale of system goes up. As a drawback, a larger view list and extended expiration time may be needed for the same level of security.

In a conclusion, Compare to Polkadot and Cosmos, our work may have larger latency but it is more decentralized. DeXTT is also a decentralized solution but our works is lighter and efficient. 

\section{CONCLUSION AND FUTURE WORKS}
This work proposes a framework for improving the scalability of multi-blockchain systems through interoperability. Under our framework, cross-chain tasks can create the externality of each system and pegged to the height of the blockchain. Since the blockchain is immutable by definition, we can use a gossip network to quickly achieve consistency among all systems. Any system that has finality and support functions in section IV can use this framework to cooperate with other systems. To our best knowledge, this is the first work to combine the security of cross-chain task with finality of all systems. Compare to some exited works, we successfully use a decentralized architecture to implement interoperability without loss of scalability or more security assumptions.

This work can be continued in many directions. First, view list could be encoded properly to reduce the burden of gossip network. Second, a more active fork detection method, like a heartbeat, could further improve the security of cross-chain tasks. Third, we did not discussed the design of proof for query message, which is also important for the efficiency of this solution. Finally, There is a high possibility to simplify the design of consensus algorithm under this framework due to the externality, which can increase the TPS of each system compare to other blockchain designs.

\begin{table}[]
	\centering
	\begin{tabular}{|l|p{0.8\linewidth}|}
		\hline
		$S$ & a system that contains several nodes and run a consensus algorithm to generate blockchain$BC_i$ with finality\\\hline
		$q$  & number of node in system \\\hline
		$r$  & minimum number of nodes in system that can make a consensus \\\hline
		$t$  & number of following blocks in system that needed to finalize a block  \\\hline
		$tx$  & transaction \\\hline
		$ctx$  & contract transaction used in CBC \\\hline
		$tx'$  & reverse transaction of $tx$ \\\hline
		$V_ij$  & the view list of $S_j$ stored by $S_i$\\\hline
		$k$  & the number of hashes in each view list $V_ij$ \\\hline
		$BC$  & the list of blocks of a system (blockchain) \\\hline
		$p_c$  & the rate of cross-chain task(ctx \& related tx) in a block \\\hline
	\end{tabular}
	\caption{The symbol used in this work.}
	\label{tab: symbols}
\end{table}

\end{document}